\documentclass[twocolumn,prl,amsmath,amssymb,showkeys]{revtex4}
\usepackage{amsmath,amssymb}
\usepackage{epsfig}
\usepackage{graphicx}
\usepackage{dcolumn}
\usepackage{bm}

\newcommand{\txtfrac}[2]{\genfrac{}{}{0pt}{1}{#1}{#2}}

\begin{document} 

\title{Replica exchange molecular dynamics simulations of amyloid peptide
aggregation}

\author{M. Cecchini}
\author{F. Rao}
\author{M. Seeber}
\author{A. Caflisch}

\email[corresponding author, tel: +41 1 635 55 21, 
fax: +41 1 635 68 62, e-mail: ]{caflisch@bioc.unizh.ch}

\affiliation{Department of Biochemistry, University of Zurich,
             Winterthurerstrasse 190, CH-8057 Zurich, Switzerland\\
             tel: +41 1 635 55 21, fax: +41 1 635 68 62,
             e-mail: caflisch@bioc.unizh.ch}

\date{\today}

\begin{abstract} The replica exchange molecular dynamics (REMD) approach is
applied to four oligomeric peptide systems.  At physiologically
relevant temperature values REMD samples conformation space and
aggregation transitions more efficiently than constant temperature
molecular dynamics (CTMD). During the aggregation process the
energetic and structural properties are essentially the same in REMD
and CTMD. A condensation stage toward disordered aggregates precedes
the $\beta$-sheet formation. Two order parameters, borrowed from
anisotropic fluid analysis, are used to monitor the aggregation
process. The order parameters do not depend on the peptide sequence
and length and therefore allow to compare the amyloidogenic propensity
of different peptides.
\end{abstract}


\maketitle

\section{Introduction}

A thorough sampling of conformational space is required to describe
the thermodynamics of complex systems such as multiple peptide chains
at finite concentrations.  Constant temperature molecular dynamics
(CTMD) techniques often fail to adequately sample conformational space
of frustrated and minimally frustrated systems which are characterized
by a rugged free-energy landscape where energy barriers between minima
are higher than the thermal energy at physiological temperature. For
this reason, a number of approaches to enhance sampling of phase space
have been
introduced~\cite{Frenkel,Berne:,Mitsutake:Generalized,
Rathore:Density}.  The parallel tempering technique (also
known as replica exchange) was developed for dealing with the slow
dynamics of disordered spin systems~\cite{Parisi:Tempering}.
%
%Replica exchange is an efficient way to simulate complex systems at
%low temperature and is the simplest and most general form of simulated
%tempering~\cite{Parisi:Tempering}.
%
Sugita and Okamoto have extended the original formulation of replica
exchange into an MD based version (REMD) and tested it on the
pentapeptide Met-enkephalin in
vacuo~\cite{Sugita:Replica}. Although in the context of fragile
liquids De Michele and Sciortino found that parallel tempering does
not increase the speed of equilibration of the (slow) configurational
degrees of freedom~\cite{Sciortino:Equilibration}, in the case of
atomistic simulations of proteins many different applications have
shown the efficiency of the method. Sanbonmatsu and Garcia have used
REMD to investigate the structure of Met-enkephalin in explicit
water~\cite{Sanbonmatsu:Structure}, and the $\alpha$-helical
stabilization by the arginine side-chain which was found to originate
from the shielding of main chain hydrogen bonds~\cite{Garcia:Alpha}.
REMD has also been applied to investigate the energy landscape of the
C-terminal $\beta$-hairpin of protein
G~\cite{Garcia:Exploring,Zhou:TheFree} and a three-helix bundle
protein~\cite{Garcia:Folding}.  REMD in implicit solvent has been used
to investigate the thermodynamics of designed 20-residue structured
peptides~\cite{Pitera:Understanding, Rao:Replica}, and recently to
study folding of a helical transmembrane protein~\cite{Im:Denovo}.

Highly ordered protein aggregates are associated with severe human
disorders including Alzheimer's disease, type II diabetes, systemic
amyloidosis, and transmissible spongiform
encephalopathies~\cite{Dobson:Proteinmis, Perutz:Glutam}. The soluble
precursors of the ordered protein deposits do not share any sequence
homology or common fold. However, X-ray diffraction data indicate a
cross-$\beta$-structure for most fibrillar
aggregates~\cite{Blake:Synchr,Malinchik:Structural}. These findings
suggest that key steps in the aggregation process may be common to all
amyloidogenic proteins.  Despite the medical relevance of amyloidoses,
many important questions about the formation of ordered aggregates
remain unanswered.  There is experimental evidence that cytotoxicity
is more pronounced for the early aggregates than for highly organized
fibrillar structures~\cite{Bucciantini:Inherent}.  Moreover, some
peptide fragments of amyloidogenic proteins display the same
properties as the full-length protein, including cooperative kinetics
of aggregation, fibril formation, binding of the dye Congo red, and
the cross-$\beta$ X-ray diffraction pattern~\cite{Balbirnie:An}. Both
findings are particularly interesting because current simulation
approaches allow significant sampling only for oligomeric peptide
systems.

There have been several lattice studies on aggregation in
proteins. These simplified models have allowed to investigate the
foldability and aggregation
propensity~\cite{Broglia:Folding,Bratko:Effect} and how interaction
potentials affect the properties of aggregation-prone
proteins~\cite{Giugliarelli:Compact}.  Harrison {\it et al.} have
shown that less stable proteins have a greater chance of assuming
alternative native states as multimers~\cite{Harrison:Thermodynamics}.
MD simulations of aggregation have been performed by using a
three-bead backbone and single-bead side chain
model~\cite{Smith:Protein}.  While this simplified model has allowed
the simulation of the competition between folding and aggregation for
two four-helix bundles it is probably not possible to extract detailed
information on energetics and sequence dependence.  Recently, a
minimalist Go model of four peptide strands~\cite{Vekhter:Modeling}
has been investigated by MD simulations in a confining sphere and the
aggregation process was shown to depend on both sequence and
environment~\cite{Friedel:Self}. Atomic models of amyloidogenic
peptides have been simulated by MD with an implicit treatment of the
solvent~\cite{Fernandez02, Gsponer:Therole, Paci:Molecular} and
explicit water
molecules~\cite{Massi:Simulation,Ma02:Molecular,Ma02,Klimov:Dissecting,
Tiana:Thermodynamics}.

%In the former, the role of complex environments on the stabilization of intermolecular
%hydrogen bonds was investigated \cite{Fernandez02}.
%The simulations of oligomers of Alzheimer's amyloid peptides
%in explicit water indicate that
%A$\beta_{16-22}$ aggregates with an antiparallel $\beta$-sheet orientation
%in agreement with solid state NMR data, and A$\beta_{16-35}$ cannot form
%linear parallel $\beta$-sheets because of unfavorable polar contacts \cite{Ma02}.
Recently, a replica exchange Monte Carlo technique has been applied to
a lattice Go model of a minimalist multichain system to study the
interplay between folding and disordered
aggregation~\cite{Bratko:Effect} but atomic model REMD applications to
ordered aggregation have not been reported yet.

In the present paper, REMD with implicit
solvent~\cite{Ferrara:Evaluation} is used to investigate the
thermodynamics of the early steps of peptide aggregation and
comparison is made with CTMD. The present work was motivated by three
questions: Is it possible to sample the early events of ordered
peptide aggregation at physiologically relevant temperatures? Do the
aggregation energetics sampled by REMD correspond to those observed in
CTMD simulations?  Are the nematic and polar order parameters,
borrowed from liquid crystal theory, useful to describe aggregation?
The simulation results indicate that all questions can be answered
affirmatively. Moreover, the ``liquid crystals'' order parameters
allow to discriminate amyloidogenic peptide sequences from those that
form only disordered aggregates.

\section{Methods}
\subsection{Model}

The MD simulations and part of the analysis of the trajectories were
performed with the CHARMM program~\cite{Brooks:CHARMM}. The oligomeric
peptide systems were modeled by explicitly considering all heavy atoms
and the hydrogen atoms bound to nitrogen or oxygen atoms (PARAM19
potential function~\cite{Brooks:CHARMM, Neria:Simulation}).  The
remaining hydrogen atoms are considered as part of the carbon atoms to
which they are covalently bound (extended atom approximation).  The
effective energy, whose negative gradient corresponds to the force
used in the dynamics, is
\begin{equation}
E({\bf r}) = E_{vacuo}({\bf r}) + G_{solv}({\bf r}) 
\label{eq:totEne}
\end{equation}
for a molecular system with atomic nuclei located at ${\bf r}=({\bf
r}_{1},...,{\bf r}_{N})$.  The PARAM19 vacuo energy function is
\begin{eqnarray}
  E_{vacuo}({\bf r}) & = & 
  \frac{1}{2}\sum_{bonds} k_b (b-b_0)^2 \; + 
  \frac{1}{2}\sum_{\txtfrac{bond}{angles}} k_{\theta} (\theta-\theta_0)^2 \nonumber \\
   & + &
  \frac{1}{2}\sum_{\txtfrac{dihedral}{angles}} k_\phi [1+\cos(n\phi-\delta)] \;\;\; \nonumber \\
   & + &
  \frac{1}{2}\sum_{\txtfrac{improper}{dihedrals}} k_\omega (\omega-\omega_0)^2  \nonumber \\
   & + &
  \sum_{i>j} \varepsilon^{\textrm{min}}_{ij} \left[ \left( \frac{d^{\textrm{min}}_{ij}}{r_{ij}} \right)^{12} - 
             2 \left( \frac{d^{\textrm{min}}_{ij}}{r_{ij}} \right)^{6} \right] \;\: \nonumber \\
   & + &
  \sum_{i>j} \frac{q_i q_j}{\epsilon(r_{ij}) r_{ij}} \nonumber
\end{eqnarray}
where $b$ is a bond length, $\theta$ a bond angle, $\phi$ a dihedral
angle, $\omega$ an improper dihedral, $r_{ij}$ is the distance between
atoms $i$ and $j$, $q_i$ and $q_j$ are partial charges, and
$d^{\textrm{min}}_{ij}$ and $\varepsilon^{\textrm{min}}_{ij}$ are the
optimal van der Waals distance and energy, respectively.  Parameters
are given in Ref.~\cite{Neria:Simulation}.

An implicit model based on the solvent accessible surface was used to
describe the main effects of the aqueous solvent on the
solute~\cite{Ferrara:Evaluation}.  In this approximation, the
solvation free energy is given by:
\begin{equation}
G_{solv}({\bf r})=\sum_{i=1}^{N} \sigma_{i}A_{i}({\bf r})
\label{eq:solv}
\end{equation}
for a molecular system having $N$ heavy atoms with Cartesian
coordinates ${\bf r}=({\bf r}_{1},...,{\bf r}_{N})$. $A_{i}({\bf r})$
is the solvent-accessible surface computed by an approximate
analytical expression~\cite{Hasel:Arapid} and using a 1.4~\AA\ probe
radius.  The solvation model contains only two $\sigma$
parameters: one for carbon and sulfur atoms ($\sigma_{C,S}=0.012$
kcal/mol {\AA}$^{2}$), and one for nitrogen and oxygen atoms
($\sigma_{N,O}$=$-$0.060 kcal/mol
{\AA}$^{2}$)~\cite{Ferrara:Evaluation}.  Hence, according to
Eq.~\ref{eq:solv} hydrophobic side chains tend to be buried within the
solute whereas hydrophilic side chains and the polar groups of the
backbone prefer to be solvent accessible.  Furthermore, ionic side
chains were neutralized~\cite{Lazaridis:Effective} and a linear
distance-dependent screening function ($\epsilon(r_{ij})=2r_{ij}$) was
used for the electrostatic interactions. The CHARMM PARAM19 default
cutoffs for long range interactions were used, i.e., a shift
function~\cite{Brooks:CHARMM} was employed with a cutoff at 7.5~\AA\
for both the electrostatic and van der Waals terms.  This cutoff
length was chosen to be consistent with the parameterization of the
force-field and implicit solvation model.  The model is not biased
toward any particular secondary structure type.  In fact, exactly the
same force field and implicit solvent model have been used recently in
MD simulations of aggregation~\cite{Gsponer:Therole, Paci:Molecular},
folding of structured peptides ($\alpha$-helices and $\beta$-sheets)
ranging in size from 15 to 31 residues~\cite{Hiltpold:Free,
Ferrara:Folding, Ferrara:Native}, and small proteins of about 60
residues~\cite{Gsponer:Molecular, Gsponer:Role}.

\subsection{REMD simulations}
The basic idea of REMD is to simulate different copies
({\it replicas}) of the system at the same time but at different
temperatures values. Each replica evolves independently by MD and
every $t_{swap}$ states $i, j$ with neighbor temperatures are swapped
(by velocity rescaling) with a probability $w_{ij}=\exp(-\Delta)$,
\cite{Sugita:Replica} where $\Delta \equiv (\beta_i - \beta_j)(E_j -
E_i )$, $\beta=1/kT$ and $ E$ is the effective energy (potential and
solvation energy, Eq.~\ref{eq:totEne}). A $t_{swap}$ of 10000 MD steps
($20\ ps$) was chosen in order to allow the kinetic and potential
energy of the system to relax.  High temperature simulation segments
facilitate the crossing of the energy barriers while the low
temperature ones explore in detail energy minima. The result of this
swapping between different temperatures is that high temperature
replicas help the low temperature ones to jump across the energy
barriers of the system.

In this study six replicas were used with temperatures (in~K): 275,
296, 319, 344, 371, 400. This range corresponds to a subset of values
used in a previous study of reversible peptide folding with the same
force-field and solvation model~\cite{Rao:Replica}. The acceptance
ratios of exchange between neighbor temperatures ranged between $15\%$
and $24\%$.  Each trajectory has a length of 2 $\mu$s for a total of
12 $\mu$s of simulation time (see Table~\ref{tab:Summ}).

\begingroup
\squeezetable
\begin{table}[t]
\begin{ruledtabular}
\caption{Simulations performed}
\label{tab:Summ}
\begin{tabular}{cccccc}
 Peptide  &  Length   &  T  & Method & IP aggregation & IA aggregation \\
 sequence & ($\mu s$) & (K) &        &     events     &      events    \\
\hline
\hline
  GNNQQNY  & $ 10 \times 0.5$ &   275   & {\sc CTMD} &  0       &  6 (19.2)~\footnote{The average time (ns) the three peptides remained aggregated in IP and IA is given in parentheses}\\
  GNNQQNY  & $ 5 \times 1.0$ &   296   & {\sc CTMD} &  3 (14.4) &  5  (1.6)\\
  GNNQQNY  & $10 \times 3.4$ &   330   & {\sc CTMD} & 54  (7.6) & 43  (1.4)\\
  GNNQQNY  & $ 2 \times 1.0$ &   371   & {\sc CTMD} &  0        &  0      \\
  GNNQQNY  & $ 6 \times 2.0$ & 275-400 & {\sc REMD} & 14 (60.3) & 15  (3.9)\\
  QQQQQQQ  & $ 6 \times 2.0$ & 275-400 & {\sc REMD} & 27 (54.8) &  2  (9.4)\\
  AAAAAAA  & $ 6 \times 1.0$ & 275-400 & {\sc REMD} &  4  (0.8) & 12  (0.9)\\
 SQNGNQQRG & $ 6 \times 2.0$ & 275-400 & {\sc REMD} &  1  (1.6) &  6  (1.0)\\
\end{tabular}
\end{ruledtabular}
\end{table}
\endgroup

\subsection{Constant temperature MD simulations}
A series of control runs were performed at constant temperature: (i)
Ten simulations at 330~K (total of 34 $\mu$s) used as a comparison for
the aggregation process between CTMD and REMD (see
Table~\ref{tab:Summ});  (ii) ten 0.5 $\mu$s simulations at
275~K and (iii) five 1 $\mu$s simulations at 296~K to compare CTMD and
REMD sampling at physiologically relevant conditions; (iv) two 1
$\mu$s simulations at 371~K to study the system near the {\it
condensation} temperature (see below).

For both REMD and CTMD, Langevin dynamics with a friction value of
$0.15 \; ps^{-1}$ was used. This friction coefficient is much smaller
than the one of water ($43 \; ps^{-1}$ at 330~K computed as $3\pi \,
\eta \, d / m$,~\cite{Hansen:Theory} where $\eta$ is the viscosity of
water at 330~K, and $d$ and $m$ are the effective diameter, i.e., 2.8
\AA \ , and mass of a water molecule, respectively) to allow for
sufficient sampling within the $\mu$s time scale of the
simulation. The small friction does not influence the thermodynamic
properties of the system.

The SHAKE algorithm~\cite{Ryckaert:Numerical} was used to fix the
length of the covalent bonds involving hydrogen atoms, which allows an
integration time step of 2 fs.  Furthermore, the nonbonded
interactions were updated every 10 dynamics steps and coordinate
frames were saved every 20 ps for a total of $5 \cdot 10^4$
conformations/$\mu$s.  A 1 $\mu$s run requires approximately 2 weeks
on a 1.4~GHz Athlon processor and the REMD simulations were run in
parallel on a Linux Beowulf cluster.

\subsection{Progress variables}
{\bf Aggregation contacts}. In-register parallel and antiparallel
aggregation contacts were defined following the prescription given in
Ref.~\cite{Gsponer:Therole}: a contact was considered to be present if
the distance between two C$_{\alpha}$ atoms placed on different
in-register strands was within 5.5 \AA. The fraction of in-register
parallel contacts $Q_{p}$ and in-register antiparallel contacts
$Q_{a}$ were used to monitor the evolution of the aggregation
process. In-register parallel and antiparallel aggregates, IP and IA
respectively, were considered formed when $Q_{p}$ and $Q_{a}$ were
larger than 0.75 ($Q_{p}, Q_{a} > 11/14$) whereas at values smaller
than 0.25 ($Q_{p}, Q_{a} < 4/14$), the system was considered
disordered. The aggregation time is defined as the temporal interval
between the first time point where $Q_{p}, Q_{a} < 0.25$ and the
following time point where $Q_{p}, Q_{a} > 0.75$.

{\bf Radius of gyration}. The radius of gyration of the oligomeric
system $R_g$ was considered to monitor the degree of {\it
condensation} and calculated using the minimum image convention. Large
values of $R_g$ indicate conformations with isolated and non
interacting peptides ({\it uncondensed phase}). Small values of $R_g$
indicate ordered as well as disordered aggregated conformations
({\it condensed phase}).

\subsection{Orientational order parameters}
\begin{figure}[h]
\includegraphics[width=8cm]{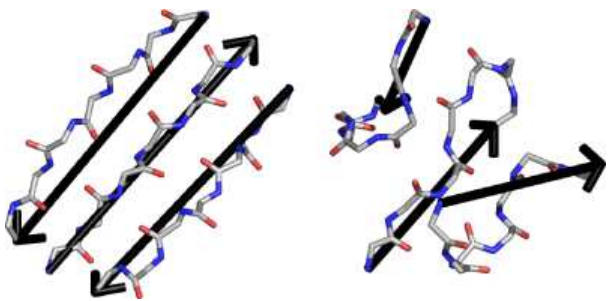}
\caption{Pictorial representation of the molecular vectors $\mathbf
{\hat z}_{i}$ (black arrows) used to compute the order parameters
$\overline{P_{1}}$ and $\overline{P_{2}}$. $\mathbf {\hat z}_{i}$
vectors are defined as full length peptide vectors (linking the
peptide's termini) and allow to clearly discriminate between ordered
(left, $\overline{P_{2}} = 0.87$) and disordered (right,
$\overline{P_{2}} = 0.46$) conformations of the system. (The pictures
were drawn using the program PyMOL~\cite{Delano:Pymol}).}
\label{fig:Image}
\end{figure}

The nematic and polar order parameters, $\overline{P_{2}}$ and
$\overline{P_{1}}$ respectively, were considered in this study. These
order parameters represent the first and second rank coefficients of
the singlet orientational distribution expanded in a Wigner
series~\cite{Rose:Elementary,Zannoni:Themolecular}, i.e., a basis set
of the Wigner rotation matrices. The nematic and polar order
parameters are widely used for studying the properties of anisotropic
fluids such as liquid crystals~\cite{Chandrasekhar:Liquid,
DeGennes:ThePhysics, Zannoni:Molecular, Berardi:CanNematic} and are
defined as
\begin{equation}
\overline{P_{2}} = \frac{1}{N} \sum_{i=1}^{N} \frac{3}{2} ({\mathbf {\hat z}_{i} } \cdot
	{\mathbf {\hat d} } )^2  - \frac{1}{2} ,
\label{eq:p2}
\end{equation}
and
\begin{equation}
\overline{P_{1}} = \frac{1}{N} \sum_{i=1}^{N} {\mathbf {\hat z}_{i} } \cdot
	{\mathbf {\hat d} },
\label{eq:p1}
\end{equation}
where $\mathbf {\hat d}$ (the director) is a unit vector defining the
preferred direction of alignment, $\mathbf {\hat z}_{i}$ is a suitably
defined molecular vector, and $N$ is the number of molecules in the
simulation box, i.e., three peptides in this study. The director is
defined as the eigenvector of the ordering
matrix~\cite{Allen:Computer} that corresponds to the largest
eigenvalue. Here, the molecular vectors $\mathbf {\hat z}_{i}$ were
defined as unit vectors linking the peptide's termini (from the N to
the C terminus, Fig.~\ref{fig:Image}). To optimally select the
$\mathbf {\hat z}_{i}$ vectors, other choices were investigated:
vectors linking the carbonyl C to the amide N of each residue
(``amide'' vectors) as well as vectors lying along the carbonyl
bonds. Similar results were obtained with the three different choices
of $\mathbf {\hat z}_{i}$. However, due to the atomic connectivity
along the backbone the ``amide'' vectors are not fully independent;
their orientations are strongly correlated and the description of the
ordered macrostates results less precise. The same is true for the
``carbonyl'' vectors. Hence, vectors linking peptide's termini were
preferred.

The order parameters (Eq.~\ref{eq:p2},~\ref{eq:p1}) change value on
going from one order macrostate to the other and should vanish when
the transition to a fully isotropic state takes place. They describe
different orientational properties of the system and yield useful and
complementary information. The nematic $\overline{P_{2}}$ describes
the orientational order of the system and discriminates between
ordered and disordered conformations. The polar $\overline{P_{1}}$
describes the polarity of the system, i.e., how much the molecular
vectors ($\mathbf {\hat z}_{i}$) point in the same direction, and
discriminates between parallel and antiparallel/mixed ordered
aggregates.

\subsection{Peptides}
To evaluate the reliability of amyloidogenic propensity estimations,
four oligomeric peptide systems were considered in this study: the
amyloid-forming heptapeptide~GNNQQNY and the soluble
nonapeptide~SQNGNQQRG both from the yeast prion Sup35 (residues 7-13
and 17-25 with the Gln/Arg mutation at position 24,
respectively)~\cite{Balbirnie:An}, the amyloidogenic poly(L-glutamine)
QQQQQQQ~\cite{Perutz:Glutamine} and the non amyloidogenic
poly(L-alanine) AAAAAAA~\cite{Perutz:Aggregation}. To reproduce the
experimental conditions~\cite{Balbirnie:An, Perutz:Glutamine,
Perutz:Aggregation}, the peptide systems derived from the yeast prion
Sup35 were modeled without blocking groups, while the Ala and Gln
repeats were both N-acetylated and C-amidated.

All simulations were performed with three peptide replicas starting
from random conformations, positions, and orientations. In the initial
random positions there was no intermolecular contact, i.e., the
peptides were separated in space. Each system was simulated in a cubic
box of 75 \AA \ per side yielding a sample concentration of 0.012
M. Since the oligomeric systems present different molecular weights,
the above reported concentration corresponds to 3.4, 3.9, 5.4 and 3.4
mg/ml for GNNQQNY, SQNGNQQRG, QQQQQQQ and AAAAAAA, respectively.

\subsection{Analysis tools}
The aggregation contacts, radius of gyration and order parameters
analysis was carried out with a GPL licensed
program~\cite{Seeber:Wordom} developed in house to manipulate and
analyze MD trajectories. The program is optimized for speed and ease
of usage so that it allows extensive processing of large amounts of
data and straightforward addition of new analysis tools. Compared to
other available programs~\cite{Brooks:CHARMM, Humphrey:VMD}, the
analysis of MD trajectories is much faster.

%
%==================================================================
%
\section{Results and Discussion}
\subsection{REMD diagnostics}
\begin{figure}[t]
\includegraphics[width=6.0cm,angle=-90]{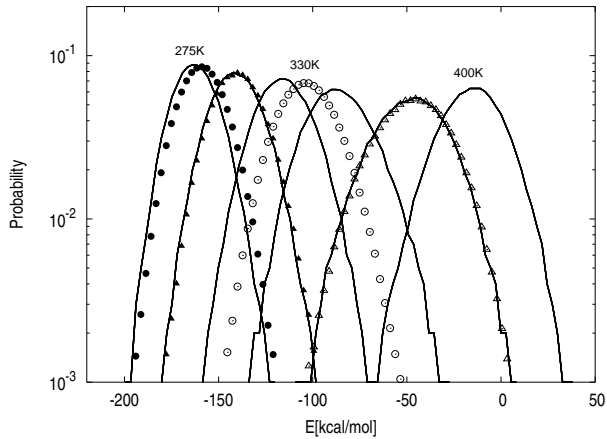}
\caption{Probability distribution of the effective energy for the REMD
(solid lines) and the CTMD control simulations (filled circles,
filled triangles, empty circles and triangles for 275, 296, 330 and
371~K, respectively).  The REMD distributions correspond to the
following temperatures (from left to right): 275, 296, 319, 344, 371,
and 400~K. The asymmetry of the curves and the temperature dependence
of the distributions indicate the presence of a phase transition
around 371~K (see text).}
\label{fig:EneHisto}
\end{figure}

\begin{figure}[t]
\includegraphics[width=7.0cm]{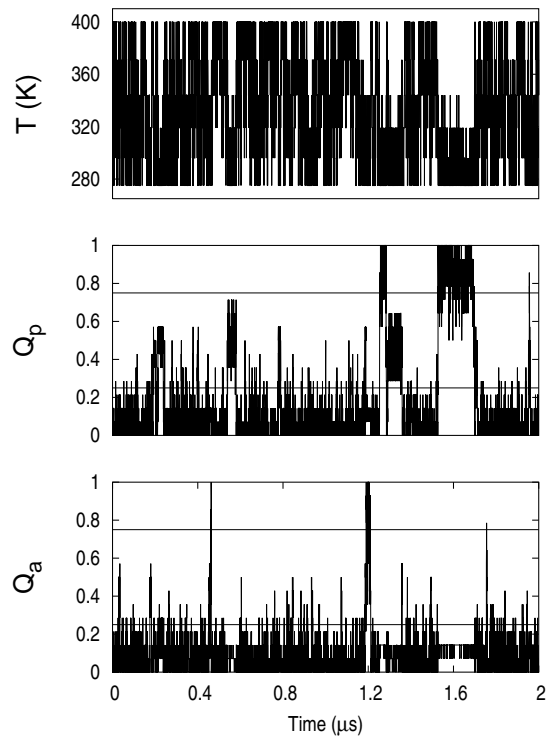}
\caption{Time series of (from top to bottom) the temperature $T$, the
fraction of in-register parallel contacts $Q_{p}$, and the fraction of
in-register antiparallel contacts $Q_{a}$ for a REMD replica. Along
the trajectory, replicas realize the desired free random walk in
temperature space (top) so that an efficient sampling of the ordered
aggregates is allowed (peaks in $Q_{p}$ and $Q_{a}$ plots). Horizontal
lines in the time series of the fraction of aggregation contacts
indicate the upper/lower thresholds used to define the ordered
aggregation/disaggregation events.}
\label{fig:TimeSeries}
\end{figure}

The set of temperatures used in a REMD simulation is crucial for a
correct and efficient sampling~\cite{Sanbonmatsu:Structure}. Since a
simple ``{\it a priori}'' protocol for selecting the optimal
temperature distribution has not been identified (yet), the choice
often follows empirical considerations~\cite{Sanbonmatsu:Structure,
Bratko:Effect, Rao:Replica}: the highest temperature of the set has to
be high enough to overcome energy barriers, while the lowest
temperature has to allow the exploration of minima.  However, given a
fixed number of replicas the temperature range cannot be too
wide. Temperature values need to be close enough to make the energy
histograms overlap (see Fig.~\ref{fig:EneHisto}) in order to guarantee
a high number of temperature swaps during a simulation run. In this
study, a set of six temperature values ranging from 275 to 400~K has
been selected (see Methods). The time series of temperature exchanges
for one of the six replicas is shown in
Fig.~\ref{fig:TimeSeries}. During the simulation, each replica visits
all the temperatures of the set several times realizing the desired
free random walk in temperature space~\cite{Sugita:Replica}.

Symbols in Fig.~\ref{fig:EneHisto} show the results from CTMD
simulations carried out at 275 (filled circles), 296 (filled
triangles), 330 (empty circles) and 371~K (empty triangles). At
330~K, the CTMD effective energy distribution is located between the
REMD distributions extracted at 319 and 344~K and shows a consistent
functional profile. At 371~K, CTMD and REMD effective energy
distributions overlap. Therefore, the energetic properties of an
aggregating system sampled by a REMD simulation at medium and high
temperatures correspond to those observed in CTMD
simulations. However, approaching the physiologically relevant
conditions the CTMD distributions tend to shift toward less favorable
energies (Fig~\ref{fig:EneHisto}, filled symbols). CTMD at low
temperature can get trapped in local energy minima and REMD is
superior in sampling conformational space~\cite{Sugita:Replica,
Rao:Replica}.

The time series of the fraction of in-register parallel contacts
($Q_{p}$) and in-register antiparallel contacts ($Q_{a}$) have been
monitored along the REMD trajectories (Fig.~\ref{fig:TimeSeries}). A
total of 14 IP and 15 IA aggregation events have been observed along
the total simulation time of 12 $\mu$s (see Table~\ref{tab:Summ}). The
average aggregation time (see Methods) was 0.74 $\mu$s for IP and 0.75
$\mu$s for IA arrangements. The average aggregation time determined
from the REMD simulation is similar to the values obtained from 34
$\mu$s CTMD simulations at 330~K. It is worth noting that in a
preliminary REMD run with higher temperatures values ($ 6 \times 1 \mu
s$, 319-465~K; data not shown) only 3 IP and 4 IA aggregation events
were sampled. The temperature range is crucial in REMD and it has to
be carefully chosen in order to speed up the conformational search of
relevant states~\cite{Fenwick:Expanded}, i.e., the {\it ordered
states} when studying aggregation. To bias the search toward
conditions where ordered states are more probable, the temperature was
set to lower values (275 to 400~K as mentioned above) and the sampling
of aggregation events turned out substantially improved.

%The average time the peptide replicas remained aggregated in the REMD
%simulation was 60.3 and 3.9 ns for IP and IA arrangements,
%respectively. These values are much larger than the ones obtained from
%CTMD simulations (see Table~\ref{tab:Summ}). The distribution of the
%lifetime of the aggregates revealed that unlike REMD, CTMD simulations
%sampled many transient aggregation events (short-lived ordered
%conformations) as well, which decrease the value of the average.

\begin{figure}[t]
\includegraphics[width=6.0cm,angle=-90]{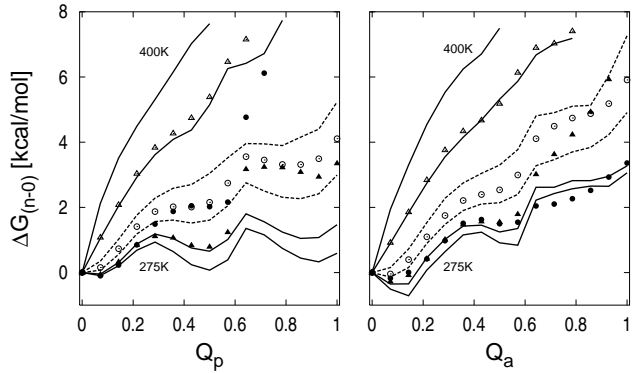}
\caption{Free-energy projections along the fraction of in-register
parallel contacts $Q_{p}$ (left) and in-register antiparallel contacts
$Q_{a}$ (right). Conformations with zero in-register contacts were
chosen as reference states. $\Delta G_{(n-0)}$ was computed as
$-k_{B}T \ln (N_{n}/N_{0})$, where $N_{n}$ indicates the number of
conformations with $n$ contacts and $k_{B}$ is the Boltzmann
constant. REMD data are shown in solid lines for all the temperature
values except for 319 and 344~K which are in dashed lines. CTMD data
are shown with symbols (filled circles, filled triangles, empty
circles and triangles for 275, 296, 330 and 371~K, respectively).}
\label{fig:Contact}
\end{figure}

Fig.~\ref{fig:Contact} shows the projections of the free energy
surface along $Q_{p}$ and $Q_{a}$ for both REMD and CTMD
trajectories. The profiles indicate that the structural properties of
the aggregating system sampled by a REMD simulation correspond to
those observed in CTMD simulations only at high and medium
temperatures. At 371~K, CTMD and REMD free energy projections
overlap. At 330~K, the CTMD free energy profiles (empty circles) are
correctly placed between REMD projections at 319 and 344~K (dashed
lines) and show patterns characterized by a well defined local minimum
at $Q_{p} > 0.7$ and a monotonic uphill trend along $Q_{a}$, fully
consistent with the profiles extracted from the REMD simulation.
However, at low temperature (275 and 296~K) the free energy
profiles extracted from CTMD and REMD trajectories are not consistent
any more and the most ``relevant'' conformations, which correspond to
in-register parallel and antiparallel arrangements ($Q_{p}, Q_{a} >
0.7$), are not correctly sampled by CTMD
(Fig~\ref{fig:Contact}, filled symbols).

\subsection{Temperature dependence of ordered amyloid peptide aggregation}
Since the energetic and structural properties of the system are not
artificially altered (see previous subsection), the REMD approach
allows to evaluate thermodynamic quantities as a function of
temperature in the chosen range~\cite{Sugita:Replica}. From the REMD
simulation performed for this study, the properties of interest
have been extracted at any temperature of the set and the aggregation
of the amyloid-forming peptide GNNQQNY has been monitored in
temperature space (275-400~K). This analysis gives interesting
insights into the amyloid aggregation process.

\begin{figure}[b]
\includegraphics[width=11.0cm, angle=-90]{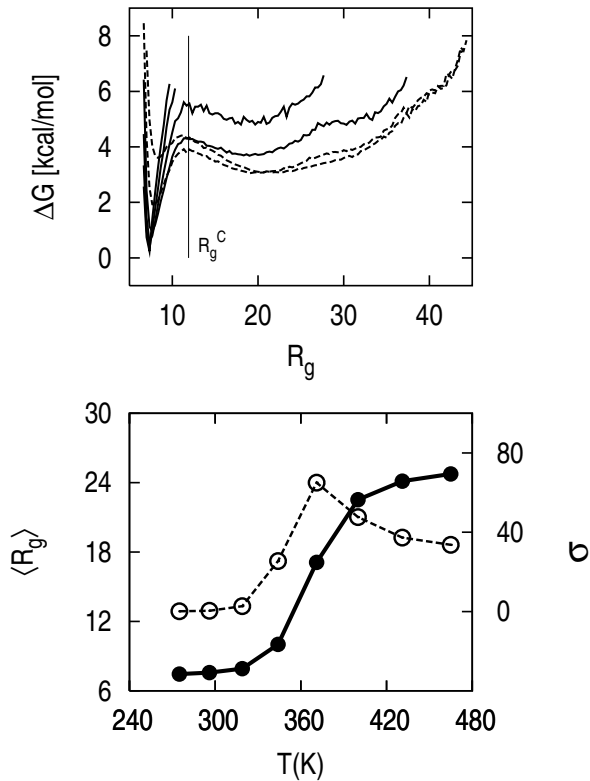}
\caption{(Top) Free-energy projections along the radius of gyration of
the oligomeric system $R_g$ computed from REMD trajectories. Solid
lines correspond to temperature values below the {\it condensation}
temperature (275-344~K); dashed lines correspond to temperature values
above the transition temperature (371 and 400~K). The lowest radius of
gyration for the {\it uncondensed} state is shown as a vertical line
(${R_{g}}^C=11.9$ \AA \ ). (Bottom) Temperature dependence of the
average radius of gyration $\left < R_g \right >$ (filled circles) and
its fluctuations $\sigma$ (empty circles). The behavior of $\left <
R_g \right >$ and $\sigma$ indicates the presence of a phase
transition around 371~K between a {\it condensed} (low T) and an
{\it uncondensed} phase (high T). Fluctuations of the radius of
gyration $\sigma$ are computed as $\left < R_g^{2} \right > - \left <
R_g \right >^2$. Data at 431 and 465~K were obtained from a
preliminary REMD run carried out in a higher temperature range ($ 6
\times 1 \mu s$, 319, 344, 371, 400, 431 and 465~K). }
\label{fig:rGyr}
\end{figure}

The effective energy histograms shown in Fig.~\ref{fig:EneHisto} are
not symmetrically distributed around their mean value and their shape
varies with temperature. The distributions, in fact, broaden toward
higher energy values at low temperature (275-344~K) and toward lower
energy values at high temperature (371-400~K). Moreover, by increasing
temperature they progressively become lower and broader till the value
of 371~K is reached. Mitsutake {\it et al.} have interpreted such a
behavior as the evidence of a phase
transition~\cite{Mitsutake:Replica-exchange}. To characterize the
transition, the radius of gyration $R_{g}$ of the oligomeric system
was considered and free energy projections along $R_{g}$ were plotted
(see Fig.~\ref{fig:rGyr}). Conformations of the system producing non
interacting peptides, namely conformations where all inter-peptide
atomic distances are larger than the long-range interactions cutoffs
(7.5 \AA \ in this case), were used to determine ${R_{g}}^C$, i.e.,
the lowest detected radius of gyration for isolated peptides (see
Fig.~\ref{fig:rGyr}). The existence of two macrostates in equilibrium
has been revealed: the first, named {\it uncondensed state}, includes
high energy conformations with one or more isolated peptides ($R_{g} >
{R_{g}}^C$ ); the second, named {\it condensed state}, consists of low
energy conformations with aggregated peptides ($R_{g}<
{R_{g}}^C$). For entropic reasons, the {\it uncondensed state} is
preferred at high temperature. By cooling down, the {\it condensed
state} is increasingly stabilized and around 371~K the fluctuations of
$R_{g}$ show a well defined peak highlighting the presence of the {\it
condensation} transition (see Fig.~\ref{fig:rGyr}). The equilibrium
between the {\it condensed} and the {\it uncondensed} macrostates is
clearly concentration dependent. If the concentration of
amyloid-forming units increases, the equilibrium is moved toward the
{\it condensed state} and the aggregation process is favored.

The free energy profiles along $Q_{p}$ and $Q_{a}$ at various
temperatures help in understanding how the nucleation process evolves
upon peptides {\it condensation}. At values of 400, 371 and 344~K both
projections show steep uphill patterns with a single free energy
minimum at $Q_{p} \approx Q_{a} \approx 0$ (see
Fig.~\ref{fig:Contact}). This means that upon {\it condensation} the
peptides are still more likely to form disordered aggregates
characterized by non-specific interactions than amyloid-forming
nuclei. In this range of temperatures, the enthalpic contribution due
to in-register backbone or side-chain interactions does not dominate
the entropic one and the growth of ordered nuclei is
forbidden. However, when the temperature decreases the entropic
contribution becomes less important and ordered in-register aggregates
start forming. As shown in Fig.~\ref{fig:Contact} in fact, below 330~K
two and one additional free energy minima appear in the projection
along $Q_{p}$ and $Q_{a}$, respectively. The observed minima
correspond to in-register parallel ($Q_{p} > 0.7$) and in-register
mixed or out-of-register ($0.3 \leq Q_{p} \leq 0.7$ and $0.4 \leq
Q_{a} \leq 0.7$) arrangements and strongly suggest that the
three-peptide system moves toward a higher degree of order when
approaching the physiologically relevant conditions.

The simulation results indicate that in the early steps of amyloid
aggregation a {\it condensation} stage toward disordered aggregates
precedes the nucleation process and the disorder-order transition, in
agreement with experimental evidence~\cite{Serio:Nucleated}.

\subsection{Disorder-order transition}
\begin{figure}[b]
\includegraphics[width=8.0cm]{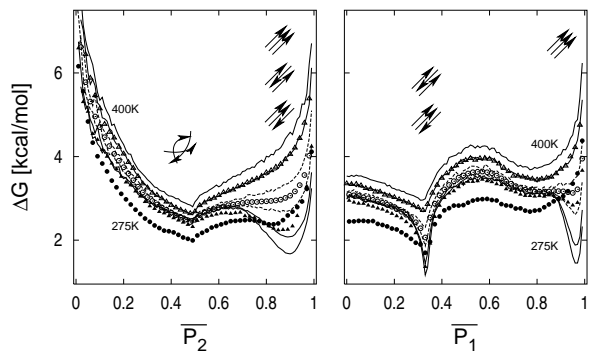}
\caption{Free-energy projections along the nematic
($\overline{P_{2}}$, left) and the polar ($\overline{P_{1}}$, right)
order parameters. REMD data are shown in solid lines for all the
temperature values except for 319 and 344~K which are in dashed
lines. CTMD data are shown with symbols (filled circles, filled
triangles, empty circles and triangles for 275, 296, 330 and 371~K,
respectively). Schematic representations of the aggregates (black
arrows) are depicted to show that order parameters yield complementary
information: $\overline{P_{2}}$ discriminates between ordered and
disordered conformations while $\overline{P_{1}}$ discriminates
between parallel and antiparallel/mixed ordered aggregates.}
\label{fig:OrdParam}
\end{figure}

In the early steps of aggregation, amyloidogenic peptides assemble
into highly ordered $\beta$-sheet structures~\cite{Balbirnie:An,
Gsponer:Therole}. During the assembly, the peptides tend to align
adopting an extended $\beta$-strand conformation and a remarkable
change in the local orientational order occurs. The aggregation of
amyloid-forming peptides may then be interpreted as an order
transition and orientational order parameters are suitable to monitor
the time evolution of the process.  Two orientational order parameters
were employed and free energy projections are shown in
Fig.~\ref{fig:OrdParam}. Along $\overline{P_{2}}$, the free energy
profiles show a first broad minimum at $\overline{P_{2}} \approx 0.5$
for any temperature of the set and a second narrower one at
$\overline{P_{2}} \approx 0.9$ for T values below 330~K. The first
corresponds to a large free energy basin where orientational order is
absent while the second corresponds to a smaller and well defined
basin with a high orientational degree of order.  Although the order
parameters should vanish when order is absent, Fig.~\ref{fig:OrdParam}
shows that this is not the case when the number of vectors is
small. Since only three peptides were simulated, a ``background''
order was always detected and the free energy minimum describing the
{\it disordered state} is placed at $\overline{P_{2}} \approx 0.5$
which is consistent with the value of $\sqrt{81 / 40 \pi \; N}$
expected for a completely randomly oriented array of $N$
molecules~\cite{Doerr:Randomness}. The order parameter
$\overline{P_{2}}$ shows the existence of two macrostates in
equilibrium: the {\it disordered state} with a high entropy content,
which corresponds to the global minimum of the free energy surface at
high temperature, and the {\it ordered state} which becomes the
global free energy minimum at low temperature. Interestingly, the free
energy profiles along $Q_{p}$ and $Q_{a}$ do not lead to the same
conclusion and the observed in-register arrangements correspond to
local minima of the free energy surface (see Fig.~\ref{fig:Contact}).

Along $\overline{P_{1}}$, two narrow and well distinct minima
corresponding to ordered macrostates at different polarity appear on
the free energy projections~(Fig.~\ref{fig:OrdParam}). The first,
displayed at $\overline{P_{1}} \approx 0.35$, describes a free energy
basin with a high-order and low-polarity content. Conversely, the
second, displayed at $\overline{P_{1}} \approx 0.95$, corresponds to a
basin with a high-order and high-polarity content. The order parameter
$\overline{P_{1}}$ discriminates between parallel and
antiparallel/mixed ordered conformations and provides complementary
information since it allows to further characterize the {\it ordered
state}. 

Symbols in Fig.~\ref{fig:OrdParam} show the free energy
projections along the order parameters from CTMD simulations. Once
again, the comparison with REMD profiles indicates that isothermal MD
(filled symbols) does not sample the ordered aggregates with their
correct statistical weight close to the physiological temperature
range.

The REMD free energy profiles along $\overline{P_{1}}$ show that at
low temperature (275 and 296~K) both polar macrostates are highly
populated. In the investigated temperature range, the system does not
show an overall polar degree and frequent jumps between ordered states
characterized by different polarity are observed. This suggests that
below the order transition the equilibrium between polar macrostates
might help amyloidogenic systems overcoming the entropy loss occurring
during nucleation. In other words the growth of amyloid-forming nuclei
might have an entropically favorable component due to the multiple
ordered macrostates.

\subsection{Sequence dependence of amyloidogenic propensity}
\begin{figure}[t]
\includegraphics[width=6.0cm,angle=-90]{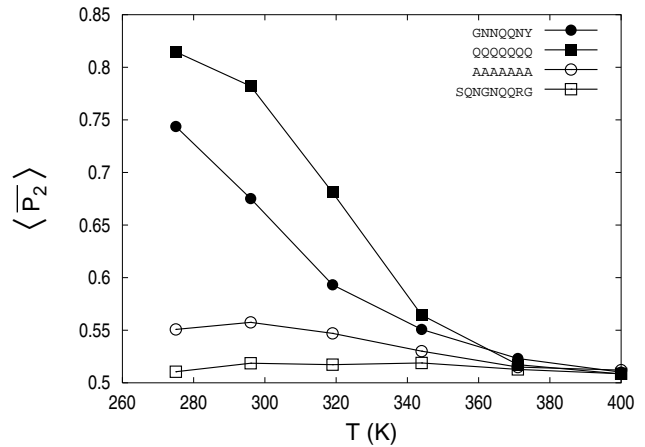}
\caption{Temperature dependence of the nematic order parameter
$\langle \overline{P_{2}} \rangle$ averaged over the canonical
ensembles sampled by REMD for four oligomeric peptide
systems. $\langle \overline{P_{2}} \rangle$ estimates the
amyloidogenic propensity of peptide systems and discriminates between
amyloidogenic (GNNQQNY and QQQQQQQ) and non amyloidogenic (SQNGNQQRG
and AAAAAAA) sequences in agreement with experimental
data~\cite{Balbirnie:An, Perutz:Glutamine, Perutz:Aggregation}.}
\label{fig:Propensity}
\end{figure}

Free energy projections along the nematic order parameter
$\overline{P_{2}}$ show how the equilibrium between the {\it ordered}
and {\it disordered state} changes in temperature space
(Fig.~\ref{fig:OrdParam}). Upon cooling, the statistical weight of the
{\it ordered state} increases and the mean of the $\overline{P_{2}}$
distribution moves toward higher values. The value of $\langle
\overline{P_{2}} \rangle$, where $\langle \cdots \rangle$ indicates
the average over the canonical ensemble, is then related to the
thermodynamic stability of the {\it ordered state} and could be used
to measure the amyloidogenic propensity of the system. $\langle
\overline{P_{2}} \rangle$ values computed at different temperatures
from REMD trajectories of the amyloid-forming peptide GNNQQNY are
shown in Fig.~\ref{fig:Propensity} with filled circles. At high
temperature, the $\langle \overline{P_{2}} \rangle$ values are close
to $0.5$ because no orientational order is present, and the system
does not show amyloidogenicity. By decreasing temperature, the
amyloidogenic propensity grows and becomes increasingly larger until
the order transition is completed. At physiologically relevant
conditions, $\langle \overline{P_{2}} \rangle \approx 0.65$ and the
system is highly amyloidogenic in agreement with experimental
data~\cite{Balbirnie:An}.

\begin{figure}[t]
\includegraphics[width=6.0cm]{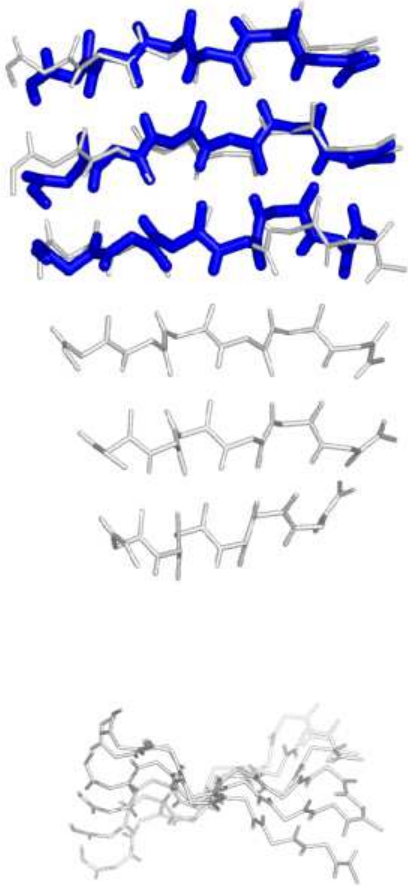}
\caption{(Top) Snapshots of ordered aggregates of three (thick
sticks) and six (thin sticks) amyloidogenic SYVIIE
peptides~\cite{Lopez:Sequence} extracted from CTMD simulations at
330~K. The simulations were performed at a sample concentration of 5
mg/ml. The overall conformation and twist of the three-stranded and
six-stranded parallel $\beta$-sheets are indistinguishable. (Bottom)
The six-stranded $\beta$-sheet upon $90^{\circ}$ rotation to better
visualize the twist. (The pictures were drawn using the program
PyMOL~\cite{Delano:Pymol}).}
\label{fig:6pep}
\end{figure}

Since the orientational order parameters do not depend on the peptide
sequence and length, the reliability of the predictions could be
further tested in sequence space. The REMD protocol was then applied
to three additional oligomeric peptide systems (see ``Methods'') and
$\langle \overline{P_{2}} \rangle$ values were evaluated to measure
and compare amyloidogenic propensities. The testing set comprises a
nonapeptide from the yeast prion Sup35 (SQNGNQQRG) experimentally
studied by Balbirnie {\it et al.}~\cite{Balbirnie:An} and two
heptapeptides (QQQQQQQ and AAAAAAA). Glutamine and alanine
homopolymers flanked by basic residues to improve solubility have been
investigated by Perutz {\it et al.}~\cite{Perutz:Glutamine,
Perutz:Aggregation}.

Experimentally, the nonapeptide SQNGNQQRG shows solubility {\it in
vivo} and {\it in vitro} and no formation of amyloid
fibrils~\cite{Balbirnie:An}. In agreement with these findings,
$\langle \overline{P_{2}} \rangle$ is smaller than $0.55$ in the whole
temperature range (Fig.~\ref{fig:Propensity}, empty squares) and the
system is considered as non-amyloidogenic.  The number of
aggregation events and the average life time of aggregation extracted
from REMD trajectories and reported in
Table~\ref{tab:Summ}. Remarkably, these quantities show that
non-amyloidogenic sequences, i.e, SQNGNQQRG and AAAAAAA, do
transiently assemble in a $\beta$-sheet conformation but still remain
soluble because their ordered aggregates do not correspond to well
defined free energy minima.

CD spectra, electron micrographs and X-ray diffraction photographs
showed that poly(L-glutamine) peptides aggregate in solution at both
pH 7.0 and 3.0 forming tightly linked $\beta$-sheets
structures~\cite{Perutz:Glutamine}. In particular, the X-ray
diffraction picture exhibits a fiber diagram of the cross-$\beta$ type
distinctive of amyloid fibrils. On the other hand, poly(L-alanine)
doesn't display amyloidogenicity and CD spectra showed
$\alpha$-helical structures at all
pHs~\cite{Perutz:Aggregation}. Again, the $\langle \overline{P_{2}}
\rangle$ patterns shown in Fig.~\ref{fig:Propensity} (filled squares
and empty circles) are consistent with experimental findings and
correctly indicate amyloidogenicity only for QQQQQQQ.

Interestingly, Fig.~\ref{fig:Propensity} allows also to compare
between amyloidogenic sequences. In fact, according to the $\langle
\overline{P_{2}} \rangle$ patterns the glutamine repeat is more
amyloidogenic than GNNQQNY at physiologically relevant conditions. To
our knowledge, no experimental data is available to verify this
finding. Testing of this prediction is a challenge for
experimentalists.

\section{Conclusions}
The present study shows that atomistic REMD simulations with implicit
solvent allow to sample the early steps of ordered aggregation of
amyloidogenic peptides at physiologically relevant temperatures.  The
free energy profiles projected along structural and orientational
progress variables are essentially the same in REMD and CTMD.  The
discrepancies at temperature values below 330~K are due to the
limitations in sampling in CTMD simulations which indicates that REMD
is a more efficient approach in the physiological range.

The early steps of amyloidosis can be interpreted as a condensation
followed by an order transition.  Therefore, the REMD simulation
results were analyzed with two order parameters originally introduced
to study liquid crystals.  Interestingly, the nematic order parameter
averaged over a canonical ensemble is able to discriminate
amyloidogenic from soluble peptides in agreement with experimental
data.

Although the present study was performed with three peptides
for reasons of computational efficiency, the description of the
ordered aggregates is likely to be independent of the size of system,
i.e., the number of simulated peptide replicas. Very recent MD
simulations of the amyloidogenic SYVIIE peptide (Cecchini {\it et
al.}, unpublished results), which has been experimentally investigated
by de la Paz and Serrano~\cite{Lopez:Sequence}, have shown ordered
aggregates of six peptides. Interestingly, the parallel $\beta$-sheet
consisting of six peptides has the same overall conformation and twist
as the three-peptide aggregate (Fig.~\ref{fig:6pep}). \\[2cm]

{\bf Acknowledgments.} We thank Dr. U.\ Haberth\"ur for running most
of the CTMD simulations and R.\ Pellarin for introducing periodic
boundary conditions in the SASA module in CHARMM (version 29). We are
grateful to E.\ Guarnera and Dr.\ E.\ Paci for helpful discussions.
We thank A. Widmer (Novartis Pharma, Basel) for providing the
molecular modeling program Wit!P which was used for visual analysis of
the trajectories.  The simulations were performed on the Matterhorn
Beowulf cluster at the Computing Center of the University of Zurich.
We thank C.\ Bollinger and Dr.\ A.\ Godknecht for setting up the
cluster and the Canton of Zurich for generous hardware support.  This
work was supported by the Swiss National Competence Center in
Structural Biology (NCCR) and the Swiss National Science Foundation
(grant no. 31-64968.01 to A.C.).

\bibliography{a-bib}

\end{document}